\documentclass[11pt,twoside]{article}
\usepackage{asp2004}
\usepackage{psfig}
\usepackage{epsf}
\usepackage{graphics}
\usepackage{lscape}
\def\edcomment#1{\iffalse\marginpar{\raggedright\sl#1\/}\else\relax\fi}
\reversemarginpar

\begin{document}
\title{The Caltech Core-Collapse Project (CCCP)}
\author{A. Gal-Yam, S. B. Cenko, D. W. Fox, D. C. Leonard, D.-S. Moon, D. J. Sand and A. M. Soderberg}
\affil{Division of Physics, Astronomy and Mathematics, California
Institute of Technology}

\begin{abstract}
The cosmological utility of type Ia Supernovae (SNe) prompted numerous 
studies of these events, and they are now well characterized observationally, 
both as individual objects and as a population. In contrast, all other types of
SNe (i.e., core-collapse events) are not as well observationally
characterized. While some
individual events have been studied in great detail (e.g., SN 1987A or
SN 1998bw), the global properties of the core-collapse SN population are little
known. However, in recent years, major drivers for change have emerged,
among them the verification of the connection between core-collapse SNe
and long-duration Gamma-Ray Bursts (GRBs), the possible utility of some 
core-collapse SNe (type II-P) as independent cosmological probes, and 
studies of core-collapse
SNe as high redshift targets for missions like SNAP and JWST. 
The Caltech Core-Collapse Project (CCCP) is a large observational program 
using the Hale 200'' and the robotic 60" telescopes at Palomar observatory to obtain 
optical photometry, spectroscopy and IR photometry of $\sim50$ nearby 
core-collapse SNe. The program is designed to provide a complete sample 
of core-collapse events, with well-defined selection criteria and uniform, 
high-quality optical/IR observations, as well as radio and X-ray light 
curves for some events. We will use this sample to characterize the 
little-studied properties of core-collapse SNe as a population. 
The sample will be used as a comparison set for studies of SNe associated 
with Gamma-Ray Bursts, to promote and calibrate the use of SNe II-P for cosmography, 
and to set the stage for investigations of SNe at high-z using coming space 
missions such as SNAP and JWST.
\end{abstract}

\section{Introduction}

For many decades Supernovae (SNe) have been a focus of scientific 
interest, as sources of heavy elements and cosmic rays, a fundamental
process shaping gas dynamics in galaxies, and, more recently, as 
cosmological distance indicators. SN explosions can be roughly
divided into two main categories: type Ia events (SNe Ia) are
believed to be thermonuclear explosions of white dwarf stars, while 
all other SNe result from core-collapse of massive progenitor stars.
The great cosmological utility of 
SNe Ia (e.g., Riess et al. 1998; Perlmutter et al. 1999;
Riess et al. 2004) spurred intensive studies of this SN sub-type.
Both detailed studies of individual nearby events, and numerous
works characterizing the properties of the SN Ia population, have been
published. This is not the case, however, for core-collapse SNe. 
While some individual core-collapse events are very well studied
(e.g., SN 1987A) the properties of the population of these events
are not well known - e.g., the average peak magnitude and light curve 
shape, the distribution around this average, the typical spectroscopic
properties and their evolution, etc.

Recently, several possible drivers for change have emerged.
The connection between hydrogen-deficient core-collapse SNe (SNe Ib/c) 
and Gamma-Ray Bursts (GRBs; Stanek et al.
2003; Hjorth et al. 2003; Matheson et al. 2003) 
resulted in great interest in these events. The possible usefulness of 
hydrogen-rich core-collapse SNe II-P
as distance estimators (e.g., Hamuy \& Pinto 2002) may offer an independent   
method to test type Ia cosmology. The need to know the average 
properties of core-collapse SNe in order to interpret the results 
from surveys for SNe
at high redshift has been demonstrated by several works 
(e.g., Dahlen \& Fransson 1999; Sullivan et al. 2000;
Gal-Yam, Maoz and Sharon 2002; Sharon 2004; Dahlen et al. 2004) and
will become even more urgent when larger planned surveys with the SNAP
and JWST space missions are carried out.

\section{Project Design and First Results}

The Caltech Core-Collapse Project (CCCP) is a large observational
program, designed to study a complete sample of nearby core-collapse
SNe, using mainly the Hale 200'' and the robotic 60'' telescopes at 
Palomar Observatory. The program will provide high quality
optical ($BVRI$) light curves, as well as multiple-epoch spectroscopy
and IR photometry, for {\it all} nearby ($R_{peak}<18.5$) core-collapse
SNe that are visible from Palomar and that have been discovered young
(no more than 30 days after explosion). We select young events using 
available photometry (e.g., SNe discovered on the rise) or
recent non-detections (i.e., SNe that are absent from images
obtained no more than 30 days prior to discovery). Young SNe are
identified as core-collapse events either through prompt spectroscopy
obtained by us (e.g., Leonard et al. 2004) or
reported by the community, or using photometric
typing methods (e.g., Poznanski et al. 2002; Gal-Yam et al. 2004a; 
Rajala et al. 2004). Including only young events in our sample removes
strong selection biases for bright or long-lasting events (such as
SNe II-P).

The compilation of well sampled optical light curves, as well as multiple
epoch spectra and IR photometry, will allow us to fully
characterize, observationally, each event in our complete sample, providing
a firm foundation for studies of core-collapse SNe as a population. CCCP
also provides data for other SN
initiatives pursued at Caltech: multiwavelength studies of SNe Ib/c and their
possible connections with GRBs (Soderberg, Kulkarni et al.); and 
testing SNe II-P as cosmological probes in effort to verify SN Ia cosmology
(Ellis and collaborators). CCCP will also support other major efforts
by the SN community, e.g., the large cycle 13 HST program to study
SNe Ia in the UV (PI Filippenko), and a similar cycle 1 GALEX program
to study core-collapse SNe in the UV (PI Gal-Yam).   

The CCCP project officially began on August 2004, and first results 
are encouraging. The robotic 60'' telescope delivers accurate, well sampled 
$BVRI$ light curves (Figure 1), while the first 200''  
runs supplied high quality spectra and NIR photometry (Figure 2).
The project is approved to continue through July 2005. More details are
available at http://www.astro.caltech.edu/$\sim$avishay/cccp.html

\begin{figure}[!ht]
\plotone{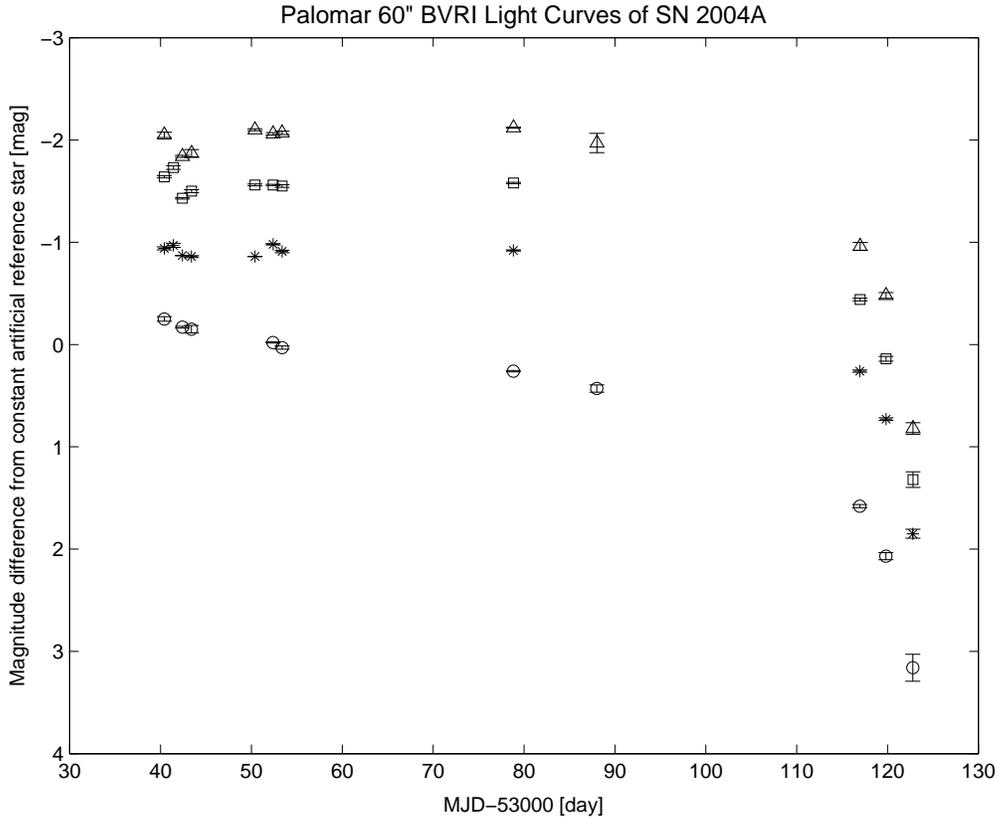}
\caption{$BVRI$ light curves of SN 2004A obtained with the robotic
60'' telescope at Palomar as part of a pilot study for CCCP. Note the
good sampling and accurate photometry obtained during the commissioning and
science verification stage of the newly roboticized telescope. Similar 
data are routinely obtained for CCCP targets starting August 2004. The
photometry was calculated using the MKDIFFLC routine (see 
http://www.astro.caltech.edu/$\sim$avishay/mkdifflc.html and
Gal-Yam et al. 2004b for additional details). }
\end{figure}

\begin{figure}[!ht]
\plotone{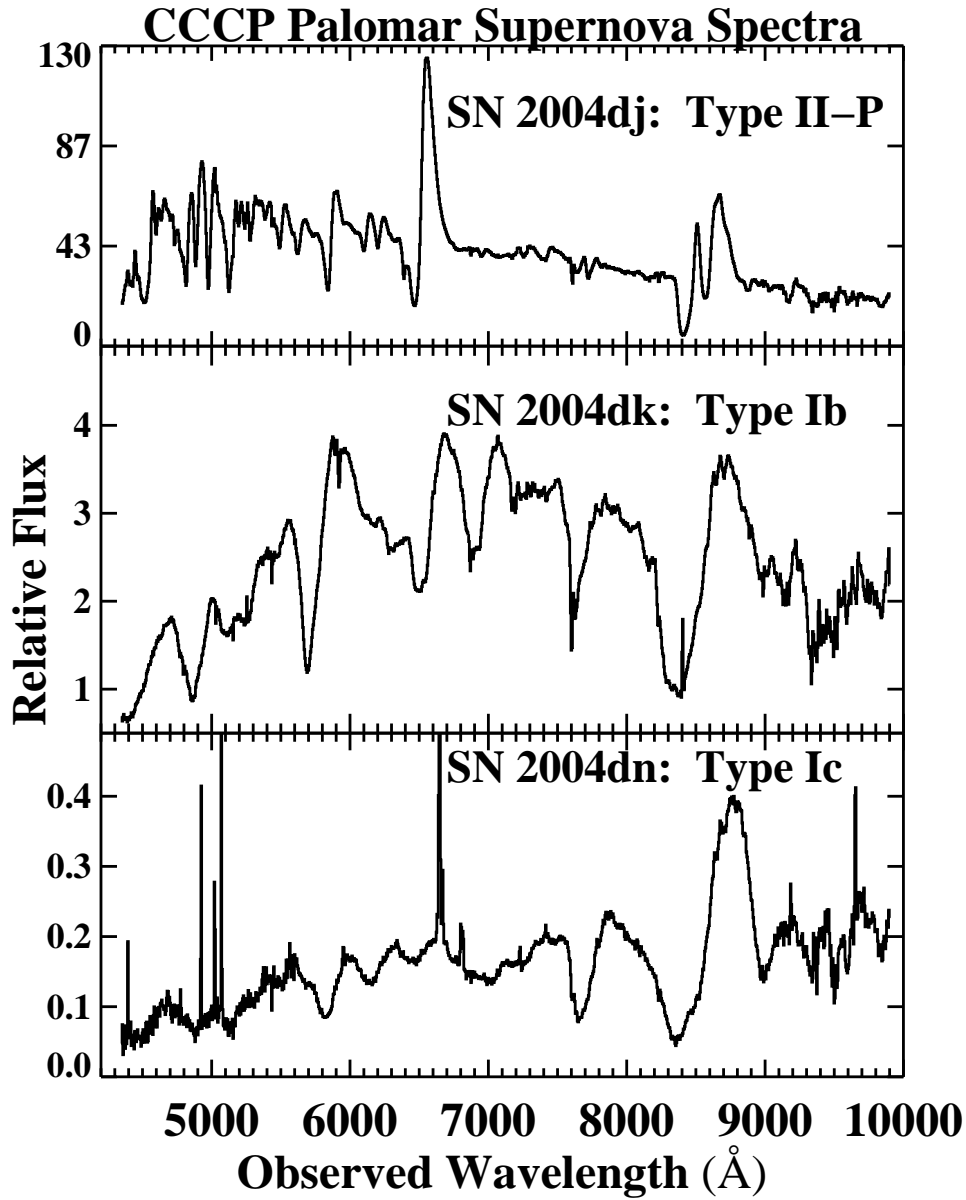}
\caption{Spectra of some of the first CCCP SNe obtained with the 200'' Hale
telescope at Palomar. A sequence of several such high quality spectra are 
obtained for each target. }
\end{figure}

\end{document}